# SanMove: Next Location Recommendation via Self-Attention Network

Huifeng Li[1], Bin Wang[1*], Sulei Zhu[1], Yanyan Xu[2*]

[1] *College of Information, Mechanical, and Electrical Engineering, Shanghai Normal University,* Shanghai, China
[2] *MoE Key Laboratory of Artificial Intelligence and AI Institute, Shanghai Jiao Tong University,* Shanghai, China
lhfqy1995@gmail.com, binwang@shnu.edu.cn, suleizhu@163.com, yanyanxu@sjtu.edu.cn

*Abstract*—Currently, next location recommendation plays a vital role in location-based social network applications and services. Although many methods have been proposed to solve this problem, three important challenges have not been well addressed so far: (1) most existing methods are based on recurrent network, which is time-consuming to train long sequences due to not allowing for full parallelism; (2) personalized preferences generally are not considered reasonably; (3) existing methods rarely systematically studied how to efficiently utilize various auxiliary information (e.g., user ID and timestamp) in trajectory data and the spatio-temporal relations among non-consecutive locations. To address the above challenges, we propose a novel method named SanMove, a self-attention network based model, to predict the next location via capturing the long- and short-term mobility patterns of users. Specifically, SanMove introduces a long-term preference learning module, and it uses a self-attention module to capture the user's long-term mobility pattern which can represent personalized location preferences of users. Meanwhile, SanMove uses a spatial-temporal guided non-invasive self-attention (STNOVA) to exploit auxiliary information to learn short-term preferences. We evaluate SanMove with two real-world datasets, and demonstrate SanMove is not only faster than the state-of-the-art RNN-based predict model but also outperforms the baselines for next location prediction.

## I. INTRODUCTION

In recent years, location-based social networks (LBSNs) like WeChat, Yelp, and Foursquare have developed rapidly. Millions of users in LBSNs have generated an abundant amount of check-in data, which enable some researches in understanding human mobility patterns at the urban scale. For example, spreading of epidemics [1], traffic congestion mitigation [2], [16] and traffic flow prediction [3], etc. Moreover, the research on next location prediction recently has attracted attention from many researchers[4], [14], [5], which can help users to explore their interesting places and help the government in urban plan.

Currently, many methods have been proposed for next location prediction, such as factorizing personalized Markov chains FPMC [6], FPMC-LR [8], and hidden Markov models (HMMs) [7]. However, these Markov-based models only consider a few observations of historical visits to learn human visit preferences and sequential transitions, which leads to poor prediction results. After that, many research efforts focus on RNN-based methods. For example, to model spatial and temporal information, Liu et al. [9] proposed the ST-RNN method to model local temporal and spatial contexts. To consider the sequential behavior of users, Feng et al. [4] designed a multi-modal RNN to capture the sequential transition. Considering different users show different dependencies on the same locations, Wu et al. [10] proposed a method named personalized long- and short-term preference learning (PLSPL) which utilize attention mechanisms and LSTM to learn the personalized preferences of users. To explore the temporal and spatial correlations between historical and recent trajectories, Sun et al. [5] proposed a context-aware long and short-term preference modeling framework to model users' preferences and a geo-dilated RNN to model the non-consecutive geographical relation between locations.

Although the above RNN-based methods have inspiring results on next place prediction, three critical problems are not well addressed: (1) RNN-based models are time-consuming to train long sequences due to its recurrent structure, and cannot be comparable with self-attention network used to capture long-dependencies [13], [18]; (2) personalized preferences generally are not considered reasonably. Existing methods for next location prediction usually cascade the embedding of user ID with the latent vector of locations in the historical and recent trajectories to capture the user's personalized preferences [10], [15] or leverage attention mechanism to capture users' personalized preferences for fixed context information [11]. However, these works lack consideration for the influence of explicit high-order interaction between users and locations. This may tend to be difficult to exploit personalized preferences effectively; (3) existing methods rarely systematically studied how to efficiently utilize various auxiliary information (user ID and timestamp) in trajectory data and the spatio-temporal relations among non-consecutive locations.

To address the above problems, we propose a novel method called SanMove that is based on the self-attention network to predict the next locations of users. Specifically, we first collect context-aware check-in locations of each user and generate the entire trajectory for each one, and then divide the entire trajectory of each user's into the historical trajectory and recent trajectory, Next, we utilize the embedding technology to embed historical trajectory and recent trajectory into the dense representation. Subsequently, we apply two based self-attention modules to capture long-term preference and short-term preference, respectively. By considering the user's representation at different timestamps as the queries of self-attention network, and representations of the locations as keys and values of self-attention network, the long-term preference learning module can capture explicit high-order interaction between users and locations and reflect the general preferences of the user. Moreover, Short-term preference can be captured by utilizing a spatial-temporal guided non-invasive self-attention (STNOVA) module which combined non-invasive self-attention (NOVA) [13] with spatio-temporal information. Finally, a concat layer and softmax layer are introduced to predict the user's next location. Our contributions are summarized as follows:

- We propose a self-attention network based sequential model to predict mobility trajectory, which consists

long-term and short-term preference learning modules and allows full parallel processing of trajectories to improve processing efficiency..

- We use a self-attention module to process with historical trajectory sequences and capture the personalized location preferences of each user. And we introduce a spatial-temporal guided non-invasive self-attention (STNOVA) module to capture sequential transitions of recent trajectories.

- We conduct extensive experiments on two check-in datasets, showing that our model improves the training speed and prediction performance compared with existing based on RNN network.

## II. THE PROPOSED METHOD

### A. Problem Formulation

In this section, we introduce the definitions and symbols used in this paper.

**Definition 1 (Trajectory).** We define a trajectory as a user's time-ordered location sequence over a period of time. Taking user $u$ as an example, $T^u = \{q_1^u, q_2^u, q_3^u, \cdots, q_k^u\}$, where each record $q_i^n \in T^u$ contains three attributes $(u, l_i, t_i)$, $u$ is the user ID, $t_i$ is the timestamp, $l_i$ is the location visited by user $u$ at time $t_i$.

**Definition 2 (Recent and Historical Trajectory).** Given a user u's trajectory $T^u$, we first split her check-in sequences into multiple sessions, $S = \{S_1, S_2, \cdots, S_n\}$, we define $T_h = \{S_1, S_2, \cdots, S_{n-1}\}$ as the user's historical trajectory, which is regarded as prior information of each user. $T_c = S_n = \{q_1^u, q_2^u, \cdots, q_k^u\}$ are the recent trajectory, which represent the location visited by the user recently.

**Problem (Next Location Prediction).** Given user u's trajectory $T_h$ and $T_c$, our goal is to recommend the next location $l_{k+1}$ by learning the historical trajectory and the recent trajectory separately.

### B. Our Model

To solve the above-defined problem, we devise a novel model SanMove. As shown in Figure 1, we first divide the whole trajectory of each user into the historical trajectory and recent trajectory. Next, we use an embedding layer to represent sparse features (e.g., $u$, $l$ and $t$), which can solve the data sparsity problem. We then use a self-attention module to capture long-term mobility preference, which can represent the user's personalized location preferences. Moreover, we use a spatial-temporal guided non-invasive self-attention (STNOVA) block to capture the short-term mobility pattern. Finally, we combine the outputs of the long-term and short-term mobility patterns to predict future trajectory.

**Trajectory Embedding Module.** To represent the user preference and spatial-temporal dependency, we jointly embed the user ID, time, and location into dense representations as the input of other modules. Specifically, for user ID $u$, we set up a trainable embedding vector $e_u \in R^d$. All user ID embeddings are denoted as a matrix $E_u \in R^{M \times d}$.

For location $l$, we utilize the same embedding method to map it into a dense vector. The embedding vector of location is represented as $e_l \in R^d$. All location embeddings are denoteds as a matrix $E_l = R^{|N+1| \times d}$. In terms of human mobility modeling, time information is also critical, therefore we also set up embeddings for the time information. Referring to [13], for each time slot $t$, we generate its embedding as follows,

$$\begin{cases} e_t(2i) = \sin(t/10000^{2i/d}) \\ e_t(2i+1) = \cos(t/10000^{2i/d}) \end{cases} \quad (1)$$

where $i$ denotes the i-th dimension. The time embedding vectors have the same dimension $d$ with the user ID embedding.

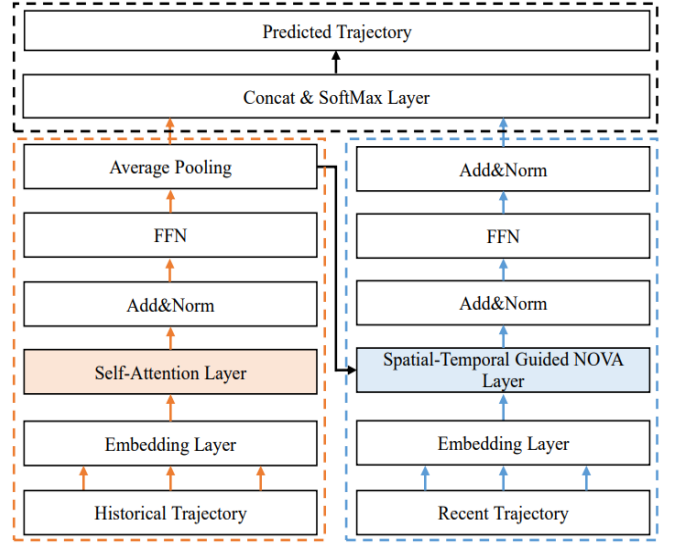

Fig. 1. The overall architecture of SanMove model.

**Long-term Preference Learning Module.** Recent years, with the great success of Transformer in machine translation, self-attention has been applied to various sequential processing tasks. Compared with the time-consuming training of RNN-based models for long sequences, the self-attention module allows fully parallel processing of data and has the ability to model long-range dependence. The self-attention module consists of a self-attention layer and a point-wise feed-forward network (FFN). The self-attention layer is defined as follows:

$$Y = Attention(XW^Q, XW^K, XW^V) \quad (2)$$

Here, $W^Q$, $W^K$, $W^V \in R^{d \times d}$ are three distinct matrices. And the basis of self-attention layer is the scaled dot-product attention, i.e.,

$$Attention(Q, K, V) = soft\max(\frac{QK^T}{\sqrt{d}})V \quad (3)$$

where $Q$, $K$, $V$ represent queries, keys, and values respectively, and they usually are the same [13]. $\sqrt{d}$ is a scale factor to avoid overly large values of the inner product. For historical trajectory $T_h$, it usually contains the user's mobility trajectory for a long time, so it can reflect the general preferences of the user's check-in behavior [10]. Therefore, in the long-term preference learning module, we first sum the time and user embedding vectors into a single one, denoted by $e_{u,t} = R^d$ as follows,

$$e_{u,t} = e_u + e_t \quad (4)$$

Then we utilize self-attention layer to compute the similarity between $e_{u,t}$ and $e_l$ to learn the importance of each location to the user at different time. where $Q = e_{u,t}$, $K = V = e_l$, and the attention layer computes a weighted sum of values in $V$, where the weight reflects the location preference of each query to keys.

To endow the model with non-linearity and encode the interactions between dimensions, we feed the output of self-attention layer into a feed forward network (FFN). Let $Y_i$ be the i-th output of of self-attention layer. Then, the feed-forward network on $Y_i$ is computed as,

$$F_i = \text{Re}\,LU(Y_iW^{(1)} + b^{(1)})W^{(2)} + b^{(2)} \quad (5)$$

where $W^{(1)}$, $W^{(2)} \in R^{d \times d}$ and $b^{(1)}$, $b^{(2)} \in R^d$. Through the transformation of self-attention module, $F_i$ aggregates specific location dependence of each user. Then, we aggregate the sequence representations via average pooling, which can reflect the general preference of the user.

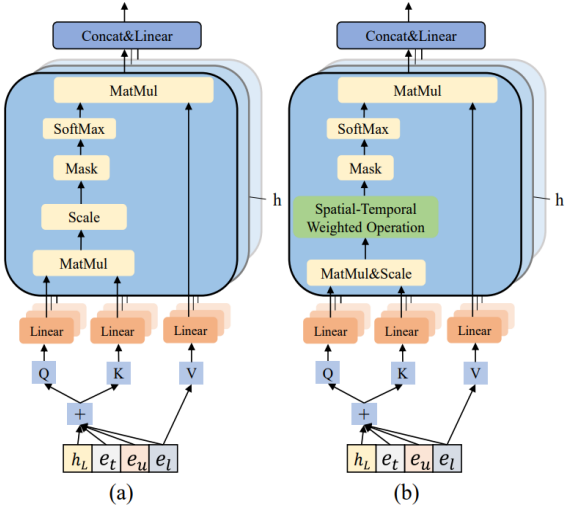

Fig. 2. (a). The architecture of the Non-invasive self-attention (NOVA); (b). The architecture of the Spatial-Temporal Guided Non-invasive self-attention (STNOVA).

**Short-term Preference Learning Module**. This part introduces a spatial-temporal guided non-invasive self-attention (STNOVA) layer and uses it to capture short-term sequential information of users. In location recommendation, existing methods leverage the mixed information of location $l$ and other auxiliary information (e.g., $u$, $t$) to learn sequence patterns of users and decode the next location [4]. However, the invasive methods have the disadvantage of compound embedding space, because location information is irreversibly fused with other auxiliary information [12]. Therefore, we utilize non-invasive self-attention (NOVA) to exploit auxiliary information to model the sequences, because it can maintain the consistency of embedding space. As shown at the (a) of Figure 2, the core idea of the NOVA is to control the information source of the self-attention components and use two sets of representations (pure location embedding and the integrated embedding) as input. Among them, the pure location embedding is expressed as $e_l = R^d$, and the integrated embedding is denoted as,

$$e_Z = e_u + e_t + e_l + h_L, \; e_Z \in R^d \quad (6)$$

where $h_L$ is the output of the long-term preference learning module and it can indicate the user's specific location preferences. Unlike NOVA in the recommender systems, we consider the associated absolute timestamps and geographical distance information which can reflect the relations among non-consecutive locations in the trajectory data [5]. Therefore, we develop a spatial-temporal guided non-invasive self-attention (STNOVA), As shown at the (b) of Figure 2, which takes into account the distance information and the time similarity information when it effectively uses auxiliary information. Specifically, we first use the integrated embedding as the value of $Q$ and $K$, and the pure location embedding as the value of $V$, and then apply them to linear transformations. Next, we input them into the scaled dot-product attention containing the spatial-temporal weighted operation. To be specific, we divide one week into 48-time slots (0-23 slots to represent hours on weekdays and 24-47 slots to represent hours on weekends) and calculate the time similarity of any two slots, as well as calculate geographical distance between locations. The definition of the STNOVA as follows:

$$\alpha_t = \frac{\exp^{\lambda_{c,j}}}{\sum_{j=0}^{47} \exp^{\lambda_{c,j}}}, \lambda_{c,j} = \frac{|T_c \cap T_j|}{|T_c \cup T_j|} \quad (7)$$

$$\alpha_s = \frac{\exp^{1/d(c,k)}}{\sum_{k=1}^{n} \exp^{1/d(c,k)}} \quad (8)$$

$$\Gamma = soft\max(\alpha_t + \alpha_s) \quad (9)$$

$$STNOVA(Q,K,V) = soft\max(\frac{\Gamma QK^T}{\sqrt{d}})V \quad (10)$$

where $T_C$ is the set of locations at current time slots $c$ and represents the locations that appear in the slot $c$. $\lambda_{c,j}$ represents the temporal similarity between the current slot $c$ and the previously visited locations' time slot $j$. $\alpha_s$ and $\alpha_t$ are weight vector between current state and previous trajectory. Same as the long-term preference learning module, we feed the output of STNOVA into an FFN to encode a non-linearity transformation following weighted summation and finally obtain the short-term sequential information of users $h_s$.

After obtaining the representations for both long-term and short-term user preferences, we make use of the softmax function to compute the probability distribution $p$ of the next location as follow:

$$p = soft\max(W_P(h_L + h_S)) \quad (11)$$

Where $W_P \in R^d$ is a trainable projection matrix. Consequently, the index of the largest probability is used as the predicted value of the next location.

| | Methods | NYC | | | | TKY | | | |
|---|---|---|---|---|---|---|---|---|---|
| | | Rec@1 | Rec@5 | NDCG@1 | NDCG@10 | Rec@1 | Rec@5 | NDCG@1 | NDCG@10 |
| Baseline | Markov | 0.1356 | 0.2732 | 0.1356 | 0.2078 | 0.1286 | 0.2500 | 0.1286 | 0.1929 |
| | LSTM | 0.1565 | 0.3189 | 0.1565 | 0.2441 | 0.1440 | 0.3051 | 0.1440 | 0.2293 |
| | DeepMove | 0.1828 | 0.3978 | 0.1828 | 0.2967 | 0.1658 | 0.3609 | 0.1658 | 0.2733 |
| | PLSPL | 0.1820 | 0.3947 | 0.1820 | 0.2948 | 0.1631 | 0.3516 | 0.1631 | 0.2615 |
| | LSTPM | 0.1883 | 0.4291 | 0.1883 | 0.3149 | 0.1773 | 0.3921 | 0.1773 | 0.2977 |
| Variants | SanMove-NOVA | 0.1941 | 0.4276 | 0.1941 | 0.3163 | 0.1778 | 0.3853 | 0.1778 | 0.2968 |
| | SanMove-P | 0.1914 | 0.4253 | 0.1914 | 0.3156 | 0.1752 | 0.3861 | 0.1752 | 0.2934 |
| | SanMove-ST | 0.1972 | 0.4301 | 0.1972 | 0.3218 | 0.1784 | 0.3949 | 0.1784 | 0.2951 |
| Proposed | **SanMove** | **0.2080** | **0.4353** | **0.2080** | **0.3279** | **0.1834** | **0.4152** | **0.1834** | **0.3087** |

TABLE I. SUMMARY OF RESULTS

## III. EXPERIMENTS

### A. Experimental Settings

**Datasets and Evaluation Metrics.** The dataset used in the experiment is Foursquare check-in datasets (NYC, TKY) which are collected from 12 April 2012 to 16 February 2013. The overall statistics of the datasets are shown in Table II. For each dataset, we begin by screening users with less than 10 records. Then, we divide each user's trajectory into multiple sub-trajectories at an interval of 72 hours, and merge the two consecutive locations if the time interval between them is less than 10 minutes. Next, we filter out sub-trajectories with less than 5 records and users with less than 5 sub-trajectories. Lastly, we use 80% of each users' trajectories as the training set and the rest as the testing set. To make fair comparisons, we adopt two evaluation metrics that are commonly-applied in previous works [17], [20], Recall@K and Normalized Discounted Cumulative Gain (NDCG@K)

TABLE II. STATISTICS OF THE EVALUATION DATASETS.

| City | # users | # locations | Timespan |
|---|---|---|---|
| New York | 1083 | 227420 | 10 months |
| Tokyo | 2293 | 573703 | 10 months |

**Baseline models and Model variants.** We will compare our model with five baseline models.

- **Markov**: modeled the latent vectors of users and locations by Matrix Factorization.
- **LSTM**: This is a variant of the RNN model which has shown effectiveness in handling sequential data.
- **DeepMove** [4]: This method learns the user's long-term preference from the history with the attention mechanism and learns short-term preference from the current trajectory using an RNN module.
- **PLSPL** [10]: A neural network model to learn the specific preference for each user, which utilizes attention mechanisms and LSTM to learn the personalized preferences.
- **LSTPM** [5]: It is the state-of-the-art model for next location prediction, which uses context-aware non-local network structure and geo-dilated RNN to capture users' long and short-term preferences respectively.

To study the contribution of the NOVA module, personalized preferences, and the spatio-temporal patterns to the overall performance of SanMove, respectively. we further compare our model with three model variants.

- **SanMove-NOVA**: a variant of SanMove uses integrated embedding as input of the self-attention components query, key, and value to model short-term preference without considering the NOVA module.
- **SanMove-P**: a variant of SanMove which doesn't consider the impact of personalized location preferences $h_L$ on short-term preference.
- **SanMove-ST**: a variant of SanMove which uses the standard NOVA network without considering the effect of spatial-temporal information.

**Parameter settings.** For baselines models' parameters, we tune it to achieve the best results or set as default values suggested by the original papers. For SanMove and its variants, we set the initial learning rate and the weight of regularization to 0.0001 and 1e-5 respectively. Meanwhile, we set the embedding size $d$ and the dimension of hidden states to 512 and use Adam optimizer to learn all the parameters. In the training process, we adopt the gradient cutting method and adjust the learning rate to guarantee the best performance of the model.

### B. Experimental Results and Analysis

In this section, we compare the performance of SanMove with baselines and variants. The performance of all methods are illustrated in Table I. From this table, we have the following observations.

First, our model SanMove outperforms the compared methods under all the metrics on the NYC and TKY datasets. Concretely, for Rec@k on the NYC dataset, our method is almost 7.2%-16.2% higher than Markov, 5.1%-11.6% higher than LSTM, 2.5%-3.7% higher than DeepMove, 2.6%-4% higher than PLSPL, and 0.6%-1.9% higher than LSTPM. For NDCG@5, our method outperforms Markov, LSTM, DeepMove, PLSPL, LSTPM, by 12.01%, 8.4%, 3.1%, 3.3%, 1.3%, respectively. On the TKY dataset, our method is also higher than other methods under all metrics.

Second, among all the methods, the Markov model's performance is the worst, indicating the powerful ability of neural network sequential models for processing trajectory data. For Rec@1 on the two datasets, the performance of the based self-attention methods is better than or similar to the state-of-the-art RNN-based model, which shows the effectiveness of self-attentive networks in predicting next locations.

Finally, for the NYC dataset, by comparing SanMove-NOVA and SanMove, we can observe that the use of the non-invasive self-attention(NOVA) to model short-term preference can make the average improvements of Rec@1 and Rec@5 on the NYC dataset are 1.38% and 0.77%, respectively. This shows that the NOVA can effectively use auxiliary information to improve the prediction ability of the network. Moreover, considering the impact of personalized location preferences $h_L$ on short-term preferences can lead to 1.66% and 1.0% improvements on average in terms of Rec@1 and Rec@5 on the NYC dataset by comparing SanMove-P and SanMove. And, by comparing SanMove-ST and SanMove, we can observe that time similarity information and distance information can affect the choice of the next location. By merging spatial-temporal information in NOVA, the average improvements of Rec@1 and Rec@5 on the NYC dataset are 1.08% and 0.52%, respectively. On the TKY dataset, the above observations are still valid.

*C. Training Efficiency Comparison*

To better understand the capacity of parallel computing of the self-attention module, we use the NYC dataset to evaluate the efficiency of SanMove and baselines. As shown in Figure 3, we record the training time cost of all methods in each epoch, and we can observe that compared with the state-of-the-art RNN-based methods, DeepMove and LSTPM, SanMove is 3.2x faster than DeepMove and is 2.4x faster than LSTPM in each epoch of training, thereby the speed of network training is improved.

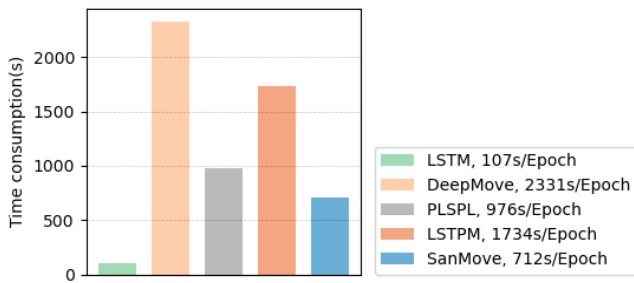

Fig. 3. Training efficiency comparison with baselines.

*D. Impact of the Embedding Size*

In this part, we investigate the sensitivity of embedding size $d$. This parameter can affect the learning ability of the model. As shown in Figure 4, on NYC and TKY datasets, we vary the embedding size value for our model from 16 to 1024, and report the performance of SanMove. We can observe that as the embedding dimension increases, the performance of our model is progressively enhanced, and when the embedding dimension is 512, the performance of our model is the best. For this reason, we choose 512 as our model's embedding dimension on two datasets.

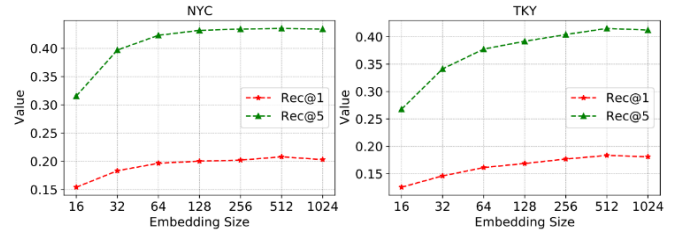

Fig. 4. The impact of embedding size on two datasets.

*E. Impact of Model Structure*

TABLE III. IMPACT OF MODEL STRUCTURE (REC@1)

| #B | NYC | TKY | #H | NYC | TKY |
|---|---|---|---|---|---|
| 1 | **0.2080** | **0.1834** | 1 | **0.2080** | 0.1778 |
| 2 | 0.2031 | 0.1761 | 4 | 0.2035 | **0.1834** |
| 3 | 0.1987 | 0.1455 | 8 | 0.2058 | 0.1821 |

In the transformer network, the self-attention module can be stacked with multiple layers and have multi-heads [14]. To study the effect of the number of self-attention module layers and the number of heads to our model, we vary the number of self-attention module layers from 1 to 3 and the number of heads from 1 to 8. The results of the evaluation are shown in Table III. From this table, we have the following observations. First of all, for the NYC and TKY datasets, 1 layer of self-attention module is better, and as the number of self-attention modules increases, the accuracy is gradually degraded. The reason may be that for trajectory datasets, deeper networks may lead to overfitting. Second, for the NYC dataset, 1 head is better, while in the TKY dataset the better number of heads is 4. This is well below the selection of 8 for the number of heads in the natural language processing task. The reason may be that the relations between locations are easier to learn than words in natural language.

## IV. CONCLUSIONS

In this paper, we propose a novel framework SanMove which is based on self-attention module, and it can predict the next location via capturing the long and short-term mobility patterns of users. In the long-term preference learning module, we use a self-attention module to capture the user's personalized location preferences which serve as the prior information for the short-term preference learning module. In the short-term preference learning module, we use a spatial-temporal guided non-invasive self-attention (STNOVA) to exploit auxiliary information to model the sequences. The extensive experimental results on two real-world datasets demonstrate the SanMove is not only faster than the state-of-the-art RNN-based predictor but also outperforms the baselines for mobility prediction.

ACKNOWLEDGMENT

This work was jointly supported by the National Key Research and Development Program of China (2020YFC2008701), the Shanghai Municipal Science and Technology Major Project (2021SHZDZX0102), and the Science and Technology Commission of Shanghai Municipality Project (2051102600).